# Zero-field topological Hall effect in BiSb/MnGa bi-layers as a signature of ground-state skyrmions at room temperature

Nguyen Huynh Duy Khang[1,2], Tuo Fan[1], and Pham Nam Hai[1,3,4*]

[1]Department of Electrical and Electronic Engineering, Tokyo Institute of Technology,

2-12-1 Ookayama, Meguro, Tokyo 152-8550, Japan

[2]Department of Physics, Ho Chi Minh City University of Education, 280 An Duong Vuong

Street, District 5, Ho Chi Minh City 738242, Vietnam

[3]Center for Spintronics Research Network (CSRN), The University of Tokyo,

7-3-1 Hongo, Bunkyo, Tokyo 113-8656, Japan

[4]CREST, Japan Science and Technology Agency,

4-1-8 Honcho, Kawaguchi, Saitama 332-0012, Japan

*Corresponding author: pham.n.ab@m.titech.ac.jp

**Abstract:** We observe the signature of zero-field ground-state skyrmions in BiSb topological insulator / MnGa bi-layers by using the topological Hall effect (THE). We observe a large critical interfacial Dzyaloshinskii-Moriya-Interaction energy ($D_\text{C}^\text{S}$ = 5.0 pJ/m) at the BiSb/MnGa interface that can be tailored by controlling the annealing temperature $T_\text{an}$ of the MnGa template. The THE was observed at room temperature even under absence of an external magnetic field, which gives the strong evidence for the existence of thermodynamically stable skyrmions in MnGa/BiSb bi-layers. Our results will give insight into the role of interfacial DMI tailored by suitable material choice and growth technique for generation of stable skyrmions at room temperature.

Skyrmions with topologically protected spin texture[1] are promising for race-track memory [2] and Brownian computing[3] applications. Thanks to their small size (~5-80 nm) [4] and low driving current density (~$10^2$ A/cm$^2$) [5], skyrmions can enable manipulation and calculation of information with ultra-low power consumption. The Dzyaloshinskii-Moriya-Interaction (DMI) is the key ingredient to stabilize the chiral domain wall of skyrmions, which gives rise to a fictitious magnetic field and can be detected by the topological Hall effect (THE).[6] Skyrmions have been observed in B20 compounds, such as MnSi, FeGe and MnN with bulk DMI originating from a break in the crystal inversion symmetry.[7-10] However bulk DMI-induced skyrmions usually exist at low temperatures. Recently, it has been shown that skyrmions can be generated at room temperature by interfacial DMI in multilayer structures consisting of heavy metal and thin ferromagnetic metal layers, such as Pt/CoFeB, Pt/Co/Ni/Co and Pt/Co.[11-15] Skyrmions observed in those multilayers are meta-stable at zero magnetic field; they can be generated by applying external perturbation such as heating,[16] bipolar pulses of magnetic field,[11] or "blowing" through a narrow channel,[17,18] and can be erased by a large magnetic field. Such meta-stable skyrmions are suitable for race-track memories. On the other hand, stable (ground-state) skyrmions, which spontaneously emerge under zero magnetic field, are suitable for Brownian computing applications. Figure 1 shows the energy landscape of meta-stable and stable skyrmions under zero magnetic field as a function of the skyrmion number $S$. Here, $S = 0,1$ correspond to the ferromagnetic (FM) state and skyrmionic (SK) state, respectively. The energy of the meta-stable SK state is higher than that of the FM state, while that of the stable SK state is lower than the FM state. So far, stable SK state has been realized in artificial skyrmion lattices defined by lithography.[19,20] For practical applications, spontaneously emerging ground-state skyrmions at room temperature are strongly required.

In this work, we demonstrate room-temperature stable skyrmions in BiSb / MnGa bi-layers. Here, BiSb is a topological insulator with high electrical conductivity[21] and strong spin Hall effect[22], and MnGa is a ferromagnet with small magnetization and perpendicular magnetic anisotropy.[23] By tailoring the interfacial DMI, we can generate meta-stable and stable skyrmions at room temperature, which were detected by the THE. In a bi-layer with a large interfacial DMI (critical interfacial DMI constant $D_S$ = 5 pJ/m), the THE was observed at room temperature even under absence of external magnetic field, indicating the existence of thermodynamically stable skyrmions. Our results give insight into the role of interfacial DMI tailored by suitable material choice and growth technique for generation of stable skyrmion at room temperature.

The BiSb/MnGa bi-layers were grown by molecular beam epitaxy on semi-insulating GaAs(001) substrates with orientation of BiSb(012) // MnGa(001) // GaAs(001). Details of the growth procedure is shown in Fig. 2(a). After a thick GaAs buffer layer was grown at 550°C, the substrate was cooled down to 50°C to grow alternative monolayers of Mn-Ga-Mn-Ga as a template.[24] In order to modulate the interfacial DMI, we annealed the template at $T_{an}$ = 350°C (sample A) or 400°C (sample B) for in 1 min to improve the MnGa surface morphology. Next, the samples were cooled down to 250°C for growth of a thick MnGa thin film with a total thickness of 5 nm. Finally, a 10 nm BiSb (Sb 15%) thin film was grown at the substrate temperature of 200°C. The growth of BiSb/MnGa bi-layers was monitored *in situ* by reflection high energy electron diffraction (RHEED). Figure 2(b)-2(d) show the RHEED pattern of the as-grown MnGa template, and after annealing at 350°C and 400°C, respectively. The RHEED pattern changes from dim to bright streaky at higher annealing temperature, indicating that the atomic ordering and morphology of the template are improved by annealing. The surface roughness of the MnGa layers grown on top of the template was evaluated by atomic force microscope (AFM), which are shown in Fig.

2(e)-2(g). The root mean square of the surface roughness $R_q$ of MnGa was reduced by annealing at higher temperatures, confirming that template annealing is effective.

To detect the signature of skyrmions in BiSb/MnGa bi-layers, we utilized the THE, which is the nondestructive electrical technique that can be incorporated to race-track memories or Brownian computer for readout of the skyrmion positions. The THE is easy to perform than using expensive tools such as the transmission X-ray holography,[25] X-ray microscopy,[26] or Lorentz force microscopy[27] with limited access. However, observation of the THE for interfacial DMI-induced skyrmions at room temperature is rare comparing with bulk DMI-induced skyrmions. The amplitude of the THE is proportional to the skyrmion density, and thus inversely proportional to the skyrmion size. The interfacial DMI-induced skyrmions are typically a few 100 nm to a few μm in diameter, for which the magneto-optical Kerr effect (MOKE) is the suitable detection technique. As shown later, the skyrmions in our BiSb/MnGa bi-layers is predicted to be 55-80 nm in diameter, thus the THE is observable while the MOKE is not. In this case, the Hall resistivity can be expressed as $\rho_{xy} = R_0 H + \rho_{AHE} + \rho_{THE}$, where the first, second and third term are the ordinary Hall effect, the anomalous Hall effect (AHE) and the THE, respectively. $\rho_{xy}$ of the BiSb/MnGa bilayers was measured by using 50 μm-wide Hall bars fabricated by conventional ultraviolet photolithography and Ar ion milling. To extract the THE resistivity from $\rho_{xy}$, we estimate AHE resistivity by measuring the magnetic circular dichroism (MCD) intensity-magnetic field (MCD-H) characteristics. MCD is a magneto optical technique that detects the difference between the reflectivity for right ($R_{\sigma+}$) and left ($R_{\sigma-}$) circularly polarized light:

$$\text{MCD} = \frac{90}{\pi} \frac{(R_{\sigma+} - R_{\sigma-})}{2R} = \frac{90}{\pi} \frac{1}{2R} \frac{dR}{dE} \Delta E,$$ where $R$ is the optical reflectivity, $E$ is the photon energy, and $\Delta E$ is the Zeeman energy of the material. Because the magnitude of MCD is proportional to

the magnetization ($\Delta E \propto M$), this measurement reflects the hysteresis of the perpendicular magnetization $M_z$.[28] Because the light wavelength used in our MCD measurements (354 nm) is much larger than the atomic scale of the chiral spin texture of skyrmions, MCD can be used to estimate AHE but not THE. Note that an external magnetic field of 8 kOe is enough to saturate the magnetization of MnGa (See the supplementary information for the MCD hysteresis loops of MnGa measured at different maximum magnetic fields). The magnetization of the MnGa layer was also measured by superconducting quantum interface devices (SQUID).

Figure 3(a) and 3(b) show the $\rho_{xy}$ (blue lines), the $\rho_{AHE}$ (red lines) estimated from MCD, and the $\rho_{THE}$ (green lines) of BiSb (10 nm) / MnGa (5 nm) bi-layers, whose MnGa template was annealed at 350°C (sample A) and 400°C (sample B) at room temperature. Two humps in the $\rho_{xy}$ curves that can be attributed to the THE were clearly observed. While the sample A shows two humps with peaks at ±1.5 kOe, the sample B shows two humps with peaks at 0 kOe. The difference between sample A and sample B can be explained by the energy landscape in Fig. 1(a) and 1(b). In sample A, the energy of the SK state is higher than the FM state at zero magnetic field (Fig. 1(a)). Thus, it is necessary to apply a non-zero magnetic field oppositely to the magnetization direction to induce skyrmions. Note that these skyrmions are meta-stable in the sense that while they can survive when the field is swept back to zero (corresponding to the bipolar pulse technique reported in Ref.11), they can be easily erased by a large magnetic field. This behavior is demonstrated in a minor loop measurement shown in Fig. 3(c) for another BiSb (10 nm) / MnGa (5 nm) bi-layer (sample C) with the MnGa template annealed at $T_{an}$ = 350°C. In contrast, the THE in sample B always re-emerges at zero magnetic field, indicating that they are stable skyrmions (Fig. 3(b)). The existence of the THE in sample B under zero external magnetic field was observed in the hysteresis curves measured at different temperatures (see the supplementary material for full

Hall resistance data at each temperature). The magnetic phase diagram of sample B as a function of the external magnetic field (*H*) and temperature (*T*) is shown in Fig. 3(d). The THE of sample B decreases with decreasing temperature and disappeared at about 125 K. The observed phase diagram of the THE is consistent with those reported for the temperature dependence of skyrmions.[27]

For comparison, we also fabricated several reference samples; a MnGa (5 nm) single layer with the MnGa template annealed at 400°C (Ref-A), a BiSb (10 nm) / MnGa (5 nm) bi-layer without the annealing step (Ref-B), and a BiSb (10 nm) / MnGa (10 nm) bi-layer with the MnGa template annealed at 400°C (Ref-C) but with much thicker MnGa layer. In all reference samples, no THE was observed (see the supplementary material for the Hall resistance data of these reference samples). The absence of the THE in the sample Ref-A indicates the observed THE in sample A-C is not due to the bulk DMI-induced skyrmions or other artifacts inside the MnGa layer, but due to interfacial DMI-induced skyrmions in the BiSb/MnGa bilayers. Meanwhile, the absence of the THE in sample Ref-B and Ref-C can be understood as below.

The existence of skyrmions is determined by competition between the Heisenberg exchange interaction, magnetic anisotropy, and DMI. The critical DMI energy for skyrmion generation is given by $D_C = 4\sqrt{AK_u}/\pi$, where $A$ is the exchange constant and $K_u$ is the magnetic anisotropy constant of MnGa. Here, the exchange constant $A$ is calculated from

$$A = (a^2 T_C \times M_s)/(2\text{g} \times \mu_B)$$

where $a$, $T_C$, $M_S$, $g$ and $\mu_B$ are the lattice constant, Curie temperature, saturation magnetization, Landé factor, and Bohr magneton constant, respectively.[29] If the DMI constant $D$ is larger than $D_C$, skyrmions can emerge. Table 1 compares $A$, $K_u$, and $D_C$ of sample B, Ref-A, Ref-B, and Ref-C. These values are estimated by measuring the out-of-plane and in-plane magnetizations of those samples (see the supplementary material for the magnetization data of these samples). Comparing sample B and sample Ref-B, one can see that the critical $D_C$ value was reduced over 2 times by annealing, mainly due to reduction of the saturation magnetization. In addition, the interfacial DMI in sample B may be larger than that of sample Ref-B because of the improved surface morphology, as can be seen in Fig. 2. Therefore, the skyrmions are observed in sample B but not in sample Ref-B. For the sample Ref-C with thicker MnGa (10 nm) thin film grown at the same condition with sample B, $D_C$ is nearly 7 times higher than that of sample B due to much larger magnetization, thus skyrmions are not observed.

In bi-layers, the critical interfacial DMI energy $D_C^S$ is related to $D_C$ as $D_C^S = D_C \times t_{MnGa}$,[15] which is 5.0 pJ/m for sample B. Table 2 compares $D_C^S$ of sample B with other multilayers reported in literature.[14,15,30] Despite the largest $D_C^S$, our BiSb/MnGa bi-layers can generate stable skyrmions. These results confirm that BiSb topological insulator is a promising material with large spin Hall angle and large interfacial DMI for generation and manipulation of skyrmions.

To further confirm the condition of $K_u$ and $D$ for generation of skyrmions in our bi-layers, we performed finite difference time domain micromagnetic simulations using the MuMax3 simulator.[31] The simulation follows steps proposed by M. Mruczkiewicz *et al.* (Ref. 32). Bloch skyrmions are assumed in the initial magnetic state, and the final magnetic state is evaluated after relaxation to the ground state under zero magnetic field. The topological number is calculated as $S = \frac{1}{4\pi} \iint m \cdot \left( \frac{\partial m}{\partial x} \times \frac{\partial m}{\partial y} \right) dxdy$ .[33] The $S$ value of the uniform ferromagnetic state, vortex state and skyrmion state are 0, 0.5, 1, respectively. The simulation parameters are the layer thickness $t = 5$ nm, saturation magnetization $M_s = 1.1 \times 10^5$ A/m, exchange constant $A = 5.4$ pJ/m, damping factor $\alpha = 0.008$, DMI constant $D = 0.5 - 2.2$ mJ/m$^2$, and $K_u = 1.0 \times 10^5 - 5.0 \times 10^5$ J/m$^3$. Here $M_s$ and $A$ value are same as those of sample B at room temperature. The simulation result for the magnetic phase diagram is showed in Fig. 4(a). The white dashed line indicates the $D_c$ value for sample B. One can see that the maximum $K_u$ at this $D_c$ for the Néel-like skyrmionic phase is around $3.2\times10^5$ J/m$^3$, which is also consistent with data reported by K. K. Meng *et al.* (Ref. 30) who showed the largest value of $K_u$ for observing the THE in MnGa/heavy metal bi-layers is approximately $3.0\times10^5$ J/m$^3$. For $K_u$ above this value, the layer changes to uniform ferromagnetic state ($S = 0$). The white dot indicates the ($D_c$, $K_u$) values for the sample B, which is within the Néel-like skyrmion phase. Thus, our observation of stable skyrmions in sample B is consistent with micromagnetic simulations. The phase diagram also suggests that the DMI constant $D$ is between $1.0$ mJ/m$^2 \leq D \leq 1.6$ mJ/m$^2$ and the corresponding interfacial DMI constant is between $5.0$ pJ/m $\leq D^S \leq 9$ pJ/m.

Figure 4(b) and 4(c) show the two dimensional and one dimensional profile of a Néel-like skyrmion simulated at $D = D_C = 1.0$ mJ/m$^2$ and $K_u = 1.2\times10^5$ J/m$^3$. The domain wall width $\Delta$ is consistent with that expected from $\Delta = \pi\sqrt{A/K_u} = 21$ nm. The size of the skyrmion is about 55 nm, which is also the lower limit of the real size of skyrmions in BiSb/MnGa bi-layers. The upper limit of the real size is 80 nm for skyrmions calculated at $D = 1.6$ mJ/m$^2$ and $K_u = 1.2\times10^5$ J/m$^3$. The observed magnitude of the THE is consistent with the estimated size of skyrmions as discussed below. The THE can be considered as the Hall effect due to a fictitious magnetic field $H_{eff}$ generated by the chiral spin texture of skyrmions with density $n_{sk}$, and each of them can be regarded as a flux quantum $\phi_0 = h/e$, where $h$ is the Plank constant and $e$ is the elementary charge. The THE resistivity is given by $\rho_H^T = PR_0H_{eff} = PR_0n_{sk}\phi_0$, [34] where $P = 42\%$ is spin polarization and $R_0 = 0.6\times10^{-5}$ μΩcm/Oe is the ordinary Hall coefficient of MnGa. Hence, we estimate that $H_{eff} \sim 7.6$ kOe and $n_{sk} = 1.1\times10^{-4}$ nm$^{-2}$, corresponding to one skyrmion per unit area of 105 nm in diameter. The estimated size of skyrmions (55-80 nm) is within the size of the unit area estimated from the observed THE resistivity. Thus, the observed THE resistivity is consistent with our micromagnetic simulation of the skyrmions in BiSb/MnGa bi-layers. Meanwhile, the unit area of 105 nm in diameter is too small to contain magnetic domains with chiral domain walls in MnGa whose size are commonly much larger than 105 nm. Furthermore, there is no observation of the THE in magnetic domains with chiral domain walls in bi-layers reported so far. Thus, the skyrmions with size of 55~80 nm in diameter are the best candidate to explain the observed large THE resistivity.

In conclusion, we have demonstrated that it is possible to generate zero-field ground-state skyrmions in BiSb topological insulator / MnGa bi-layers. Despite a large critical interfacial DMI energy ($D_\text{C}^\text{S}$ = 5.0 pJ/m) at the BiSb/MnGa interface, stable skyrmions were observed in the sample with improved morphology at room temperature under zero magnetic field. Our results give insight into the role of interfacial DMI tailored by suitable material choice and growth technique for generation of stable skyrmions.

Note added. Recently, we noticed a similar result of the zero-field ground-state skyrmions in Fe/Gd multilayers reported by another group, [35] where dipolar interactions rather than DMI are utilized to stabilize skyrmions.

See supplementary material for (S1) AHE and MCD of sample B at various temperature, (S2) Hall resistance and (S3) the out-of-plane and in-plane magnetization curves of reference sample Ref-A, Ref-B, and Ref-C, and (S4) The MCD-$H$ hysteresis loops of sample B at different maximum external magnetic fields.

## Acknowledgement

This work is supported by JST-CREST (JPMJCR18T5). The authors thank M. Tanaka Laboratory at the University of Tokyo and S. Nakagawa Laboratory at Tokyo Institute of Technology for their supports and helps of MCD, AFM and SQUID measurements.

**Figures and Captions**

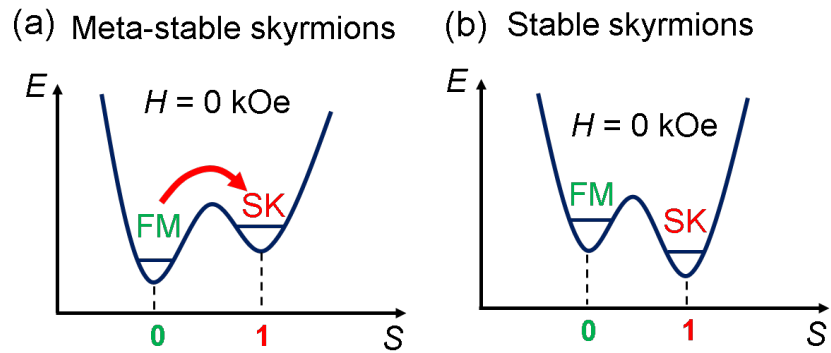

**Fig. 1.** Energy landscape for **(a)** meta-stable skyrmions and **(b)** stable skyrmions under zero magnetic field. Here, the skyrmion number $S = 0,1$ represent the ferromagnetic (FM) state and the skyrmionic (SK) state, respectively. In **(a)**, the meta-stable skyrmions can be formed by external perturbation, while in **(b)** the skyrmions form spontaneously.

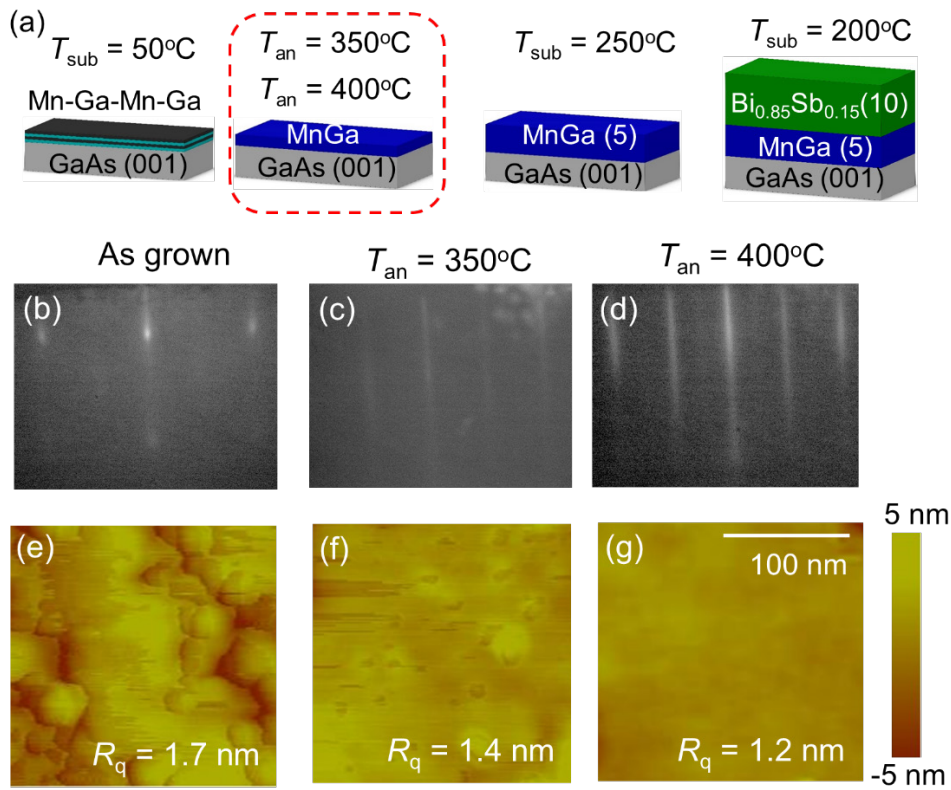

**Fig. 2. (a)** Schematic illustration of the growth process. An annealing step of the MnGa template was employed to tailor the interfacial DMI. **(b-d)** RHEED patterns observed along the GaAs[$\bar{1}10$] direction for the as-grown MnGa template, and after annealing at 350°C and 400°C, respectively. **(e-g)** Surface morphology of the MnGa thin films measured by atomic force microscopy.

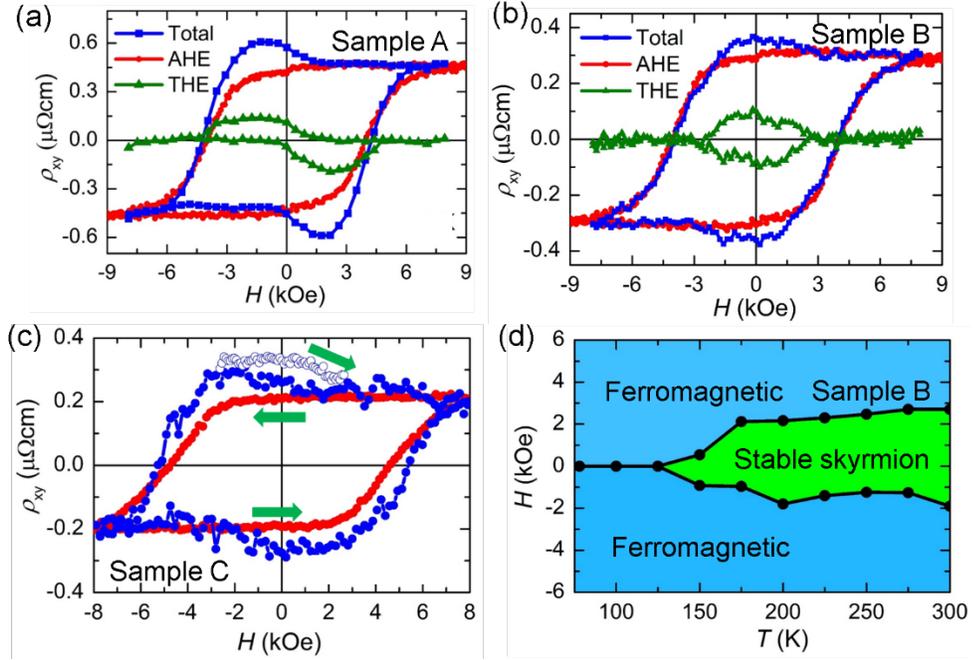

**Fig. 3. (a-b)** Room-temperature THE resistivity (green) extracted from the total Hall effect resistivity (blue) and AHE resistivity (red) derived from magnetic circular dichroism intensity-magnetic field measurements of BiSb(10 nm) / MnGa(5 nm) bi-layers, whose MnGa template was annealed at 350°C (sample A) and 400°C (sample B), respectively. **(c)** Major loop (solid blue circles) and minor loop (open blue circles) of the total Hall effect in another BiSb (10 nm) / MnGa (5 nm) bi-layer (sample C) with the MnGa template annealed at 350°C. **(d)** Magnetic phase diagram of sample B as a function of the external magnetic field ($H$) and temperature ($T$).

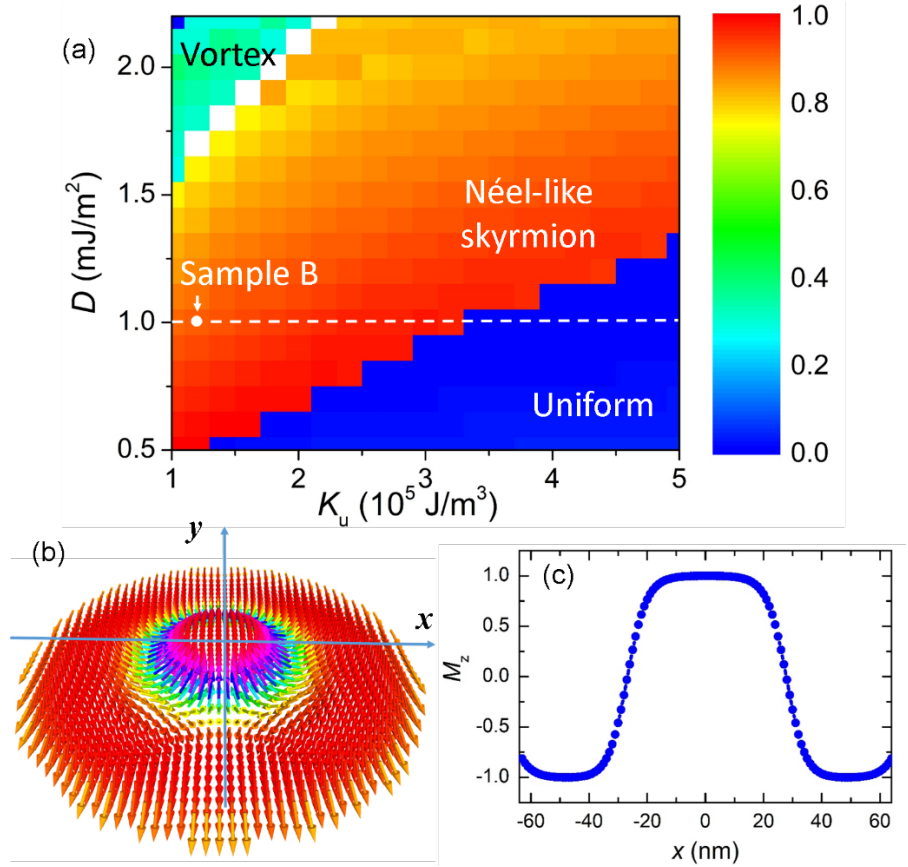

**Fig. 4. (a)** The skyrmion number $S$ as a function of magnetic anisotropy constant $K_u$ and DMI constant $D$. (b) Néel-like skyrmion simulated with $K_u = 1.2 \times 10^5$ J/m$^3$ and $D = 1.0$ mJ/m$^2$. **(b)** $M_z$ profile of the Néel-like skyrmion along the $x$ direction. The size of the skyrmion is about 55 nm.

**Table 1.** Comparison of the saturation magnetization $M_s$, exchange constant $A$, anisotropy constant $K_u$, and critical DMI energy $D_c$ for sample B and several reference samples. The number in the parenthesis after each layer is the thickness in nm unit.

| Sample | Stack | Template annealing | $M_s$ ($10^5$ A/m) | $A$ (pJ/m) | $K_u$ ($10^5$ J/m³) | $D_c$ (mJ/m²) | THE |
|---|---|---|---|---|---|---|---|
| B | BiSb(10)/MnGa(5) | Yes | 1.1 | 5.4 | 1.2 | 1.0 | Yes |
| Ref-A | MnGa(5) | Yes | 2.2 | 10.9 | 1.3 | - | No |
| Ref-B | BiSb(10)/MnGa(5) | No | 2.4 | 11.7 | 3.1 | 2.4 | No |
| Ref-C | BiSb(10)/MnGa(10) | Yes | 4.0 | 19.6 | 15.2 | 7.0 | No |

**Table 2.** Comparison of the critical interfacial DMI energy $D_C^S$ observed in this work and other bi-layers reported so far.

| Bi-layer | $D_C^S$ (pJ/m) |
|---|---|
| Pt/CoFeB (Ref.14) | 0.8 |
| Pt/Co (Ref. 15) | 1.37-2.17 |
| Pt or Ta /MnGa (Ref. 30) | 2.60-4.32 |
| **BiSb/MnGa (this work)** | **5.0** |

# Zero-field topological Hall effect in BiSb/MnGa bi-layers as a signature of ground-state skyrmions at room temperature

Nguyen Huynh Duy Khang[1,2], Tuo Fan[1], and Pham Nam Hai[1,3,4*]

[1]Department of Electrical and Electronic Engineering, Tokyo Institute of Technology,

2-12-1 Ookayama, Meguro, Tokyo 152-8550, Japan

[2]Department of Physics, Ho Chi Minh City University of Education, 280 An Duong Vuong

Street, District 5, Ho Chi Minh City 738242, Vietnam

[3]Center for Spintronics Research Network (CSRN), The University of Tokyo,

7-3-1 Hongo, Bunkyo, Tokyo 113-8656, Japan

[4]CREST, Japan Science and Technology Agency,

4-1-8 Honcho, Kawaguchi, Saitama 332-0012, Japan

*Corresponding author: pham.n.ab@m.titech.ac.jp

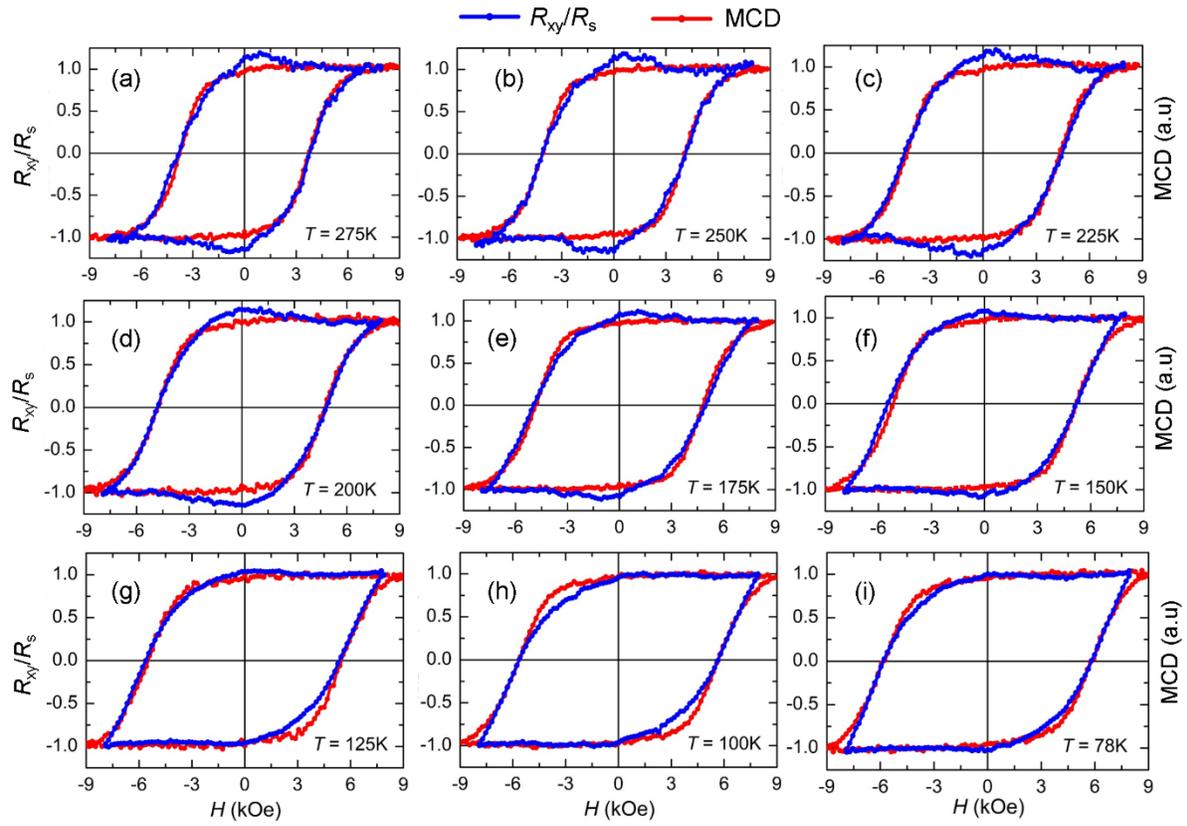

**Fig. S1.** Hall resistance normalized by its maximum value (blue) and magnetic circular dichroism intensity (red) hysteresis of sample B at various temperatures.

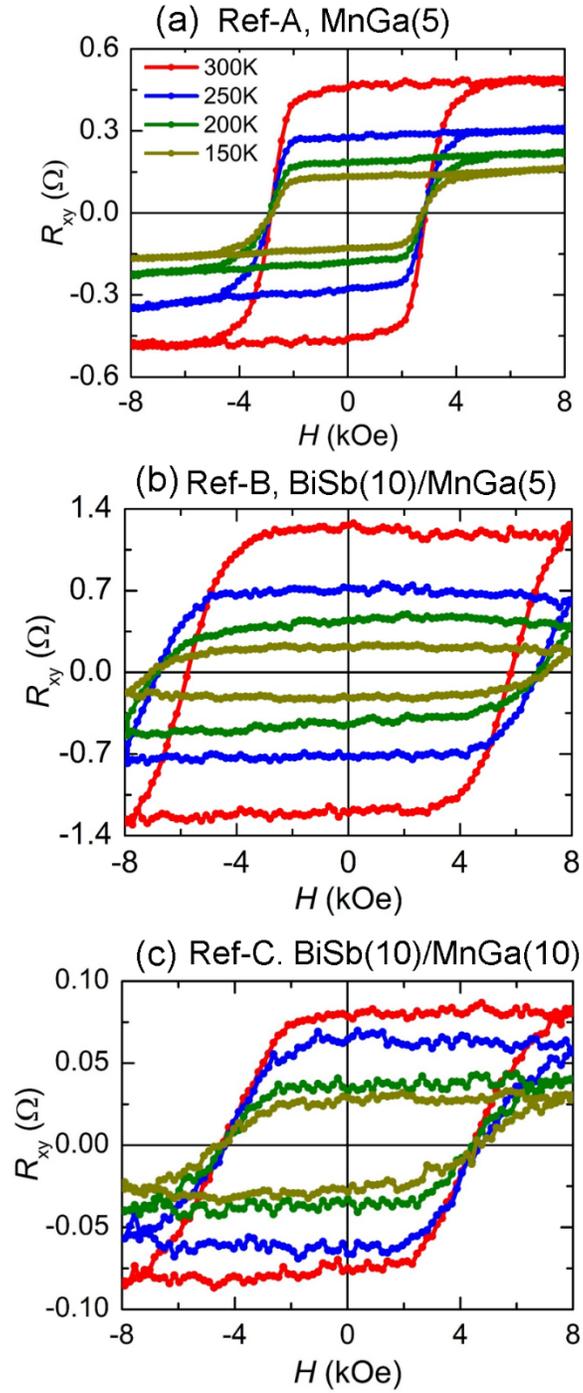

**Fig. S2.** Hall resistance of several reference samples. **(a)** the MnGa (5 nm) single layer with the template annealed at 400°C (Ref-A), **(b)** the BiSb (10 nm) / MnGa (5 nm) bi-layer without the annealing step (Ref-B), and **(c)** the BiSb (10 nm) / MnGa (10 nm) bi-layer with the template annealed at 400°C (Ref-C).

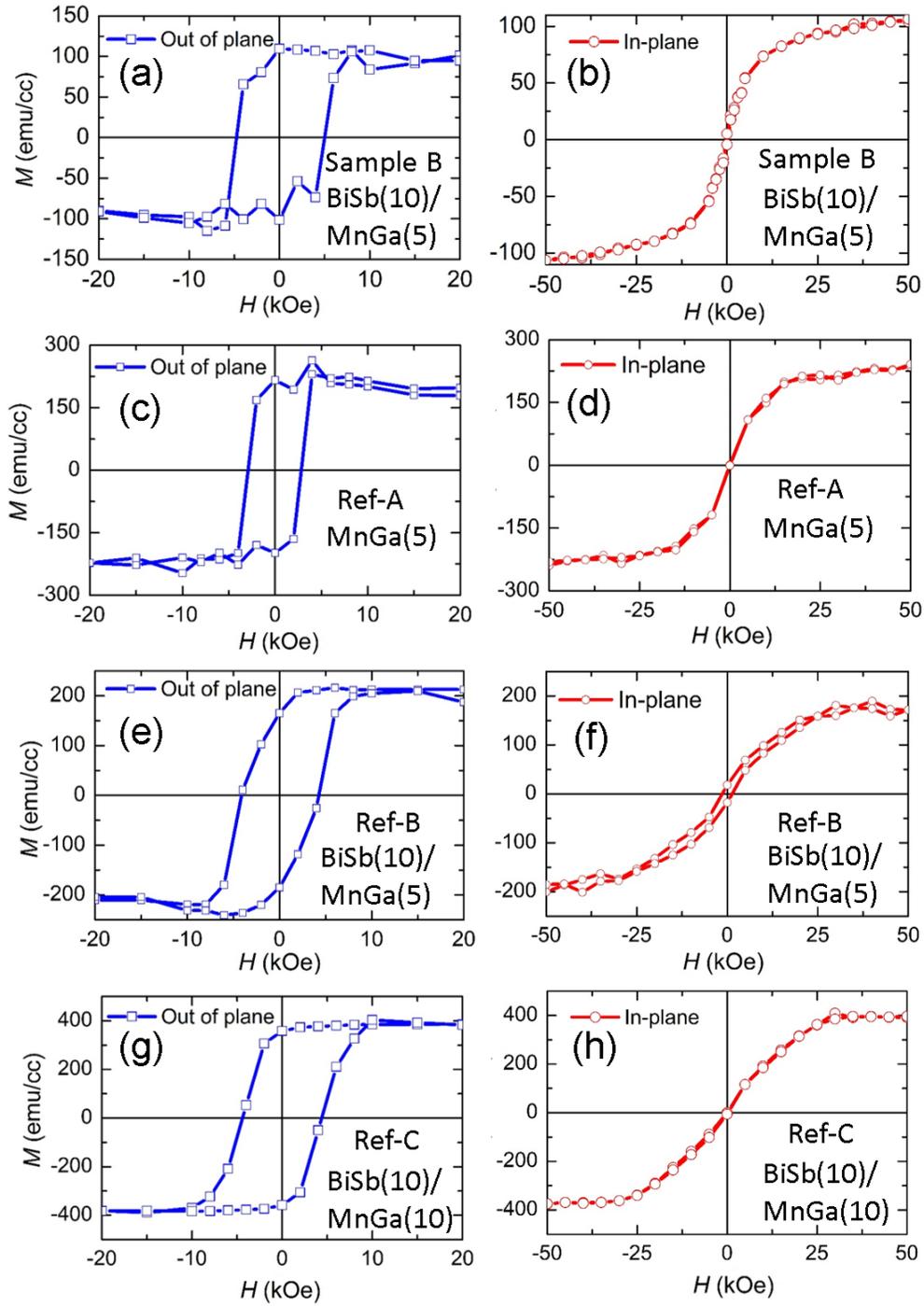

**Fig. S3.** The out-of-plane and in-plane magnetization curves of **(a)(b)** sample B, **(c)(d)** Ref-A, **(e)(f)** Ref-B, and **(g)(h)** Ref-C. The number in the parenthesis is the thickness in nm unit.

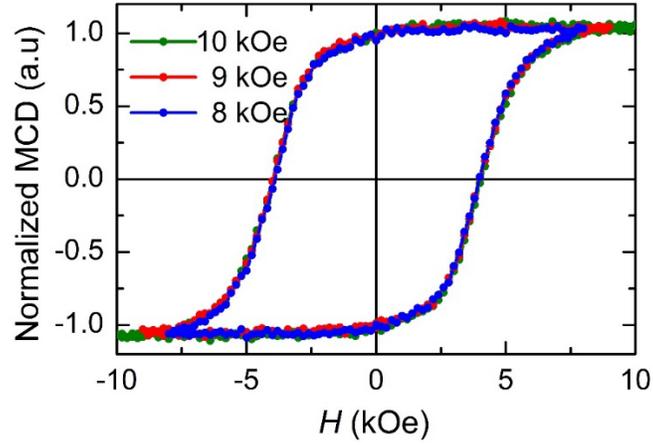

**Fig. S4.** The MCD-$H$ hysteresis loops of sample B (BiSb(10 nm) / MnGa(5 nm) bi-layer), whose MnGa template was annealed at 400°C, measured at different maximum magnetic field $H_{max}$ of 10 kOe (green), 9 kOe (red), and 8 kOe (blue).